\documentstyle[12pt,twoside]{article}

\newcommand{\tha}{{\vartheta}_1}
\newcommand{\thh}{{\vartheta}_4} 
\newcommand{\be}{\begin{equation}}
\newcommand{\ee}{\end{equation}}
\newcommand{\beqs}{\begin{eqnarray}}
\newcommand{\eeqs}{\end{eqnarray}}

\newcommand{\tr}{{\rm tr}}

\newcommand{\half}{{1 \over 2}}

\newcommand{\bc}{{\rm c}}
\newcommand{\n}{{\rm n}}
\newcommand{\m}{{\rm m}}

\def\PL{{\it Phys. Lett.\ }}

\def\PR{{\it Phys. Rev.\ }}
\def\PRL{{\it Phys. Rev. Lett.\ }}

\def\JMP{{\it J. Math. Phys.\ }}

\def\LNC{{\it Lett. Nuovo Cimento \ }}

\expandafter\ifx\csname mathbbm\endcsname\relax

\else

\fi
\begin{document}
\begin{titlepage}
\begin{flushleft}  
       \hfill                       RU-01-19-B\\
       \hfill                       December 2001\\
\end{flushleft}
\vspace*{3mm}
\begin{center}
{\LARGE Calogero-Moser models with noncommutative
spin interactions \\}
\vspace*{12mm}
\large Alexios P. Polychronakos\footnote{On leave from Theoretical
Physics Dept., Uppsala, Sweden; E-mail: poly@teorfys.uu.se} \\
\vspace*{5mm}
{\em Physics Department, Rockefeller University \\
New York, NY 10021, USA \/}\\
\vspace*{4mm}
and\\
\vspace*{4mm}
{\em Physics Department, University of Ioannina \\
45110 Ioannina, Greece\/}\\
\vspace*{15mm}
\end{center}

\begin{abstract}
We construct integrable generalizations of the elliptic
Calogero-Sutherland-Moser model of particles with spin,
involving noncommutative spin interactions. The spin
coupling potential is a modular function and, generically,
breaks the global spin symmetry of the model down to
a product of $U(1)$ phase symmetries. Previously known
models are recovered as special cases.

\end{abstract}

\vspace*{10mm}
PACS: 03.65.Fd, 71.10.Pm, 11.10.Lm, 03.20.+i

\end{titlepage}

\section{Introduction} 

The inverse-square interacting particle system \cite{Cal,Suth,Mos} 
and its spin generalizations \cite{GH,Woj,HH,Kaw,MP1,HW} are important 
models of many-body systems, due to their exact solvability and
intimate connection to spin chain systems \cite{Hal,Sha,FM,APsc,BHUW},
2-dimensional Yang-Mills theories \cite{GN,MP2,LSK} etc.
Reference \cite{OP} is a classic review, while \cite{APhouches,DHP,BCS}
present newer different perspectives.

The prototype of these models is the spin-Calogero (`rational') 
scattering model of particles on the line carrying $U(\n)$ spin
and interacting with two-body inverse-square potentials with
a $U(\n)$-invariant spin coupling. Most other models 
can be obtained as appropriate reductions of this model, taking
advantage of its discrete or continuous symmetries \cite{APred}.
In particular, generalizations involving $U(\n)$ non-invariant
interactions can be obtained this way, recovering the trigonometric
models derived in \cite{BL,APmmm} and extending them to the
elliptic case \cite{APred}. 

An unrelated development has been the recent progress in noncommutative
field theory and matrix models. Spatial noncommutativity can be traced
back to Heisenberg and naturally
arises in lowest Landau level physics \cite{RJ}. Its current manifestation
originates in matrix, string and membrane theory \cite{BFSS,SW} and 
came into focus with the work of Connes, Douglas and Schwartz \cite{CDS}.

So far these two fields remained unrelated. In this letter we show how
they can be cross-fertilized by borrowing notions of noncommutative
geometry and applying them in the reduction scheme of the Calogero
model to obtain a new
integrable elliptic model involving non-$U(\n)$ invariant noncommutative
spin interactions. Such a modification
of the spin interaction may serve to test the `flavor stiffness' of the original
spin model, to stress the degeneracy structure of the energy spectrum and
to identify universality features of this class of models. 

\section{The reduction scheme}

The basic technique that we will use consists of reducing 
a system of infinitely many particles with spin to a finite system with 
generalized interactions. The reader should
refer to \cite{APred}, and especially to the elliptic case with spin,
for a more detailed description of the method.

The starting point is the spin-Calogero system with classical 
$U(\n)$ degrees of freedom. This system can be obtained,
for instance, from the model in \cite{GH,Woj} (which can
itself be obtained as a reduction of a hermitian matrix model
\cite{KKS} into nontrivial angular momentum sectors) 
by redistributing the global $U(N)$ degrees of freedom of this model
into individual particle spins, or, alternatively, from the
infinite-volume classical limit of the spin model derived and solved
in \cite{MP2}. The hamiltonian for $N$ particles reads
\be
H = \sum_{i=1}^N \half p_i^2 + \half \sum_{i \neq j} {\tr(S_i S_j )
\over x_{ij}^2}
\label{Hcal}
\ee
$x_i$ and $p_i$ are one-dimensional canonical coordinates and momenta; 
$S_i$ are a set of independent classical $U(\n)$ spins of rank one and length
$\ell$, that is, $\n \times \n$ rank-one hermitian matrices satisfying
\be
\tr (S_i )^2 =  \ell^2
\label{lengthl}
\ee
and with Poisson brackets
\be 
\{ (S_i)_{ab} , (S_j)_{cd} \} = -i \delta_{ij}
\left[ (S_i)_{ad} \, \delta_{cb} - \delta_{ad} \, (S_i)_{cb} \right]
\ee
Such spins can be realized in terms of oscillators \cite{MP2}: 
\be
(S_i )_{ab} = {\bar A}_i^a A_i^b
~,~~ a,b=0 \dots \n-1
\label{Sa}
\ee
where $( A_i^a \,,\, {\bar A}_i^a)$ are a set of $\n N$ independent
classical harmonic oscillator canonical pairs with Poisson brackets:
\be
\{ A_i^a , {\bar A}_j^b \} = i \delta_{ij} \, \delta_{ab}
\ee
and satisfying the constraint
\be
\sum_a {\bar A}_i^a A_i^a = \ell ~~{\rm for~all}~i
\label{Acon}
\ee

In the above model we can analytically continue the coordinates
$x_i$ and momenta $p_i = {\dot x}_i$ to the complex plane.
The integrability and solvability of this model trivially extends
to the complex case. Such extensions will be useful, provided
that we can identify a real subsystem, which will be the physical
system of interest. 

The hamiltonian $H$ is invariant under particle permutations,
coordinate space translations and global spin rotations:
\be
i \to p_{_N} (i) ~;~~~ x_i \to x_i +\bc ~;~~~ S_i \to U S_i U^{-1}
\ee
where $p_{_N}$ is an element of the permutation group $S_N$, $\bc$ is
a constant complex parameter and $U$ is a constant $U(\n)$ matrix.
We can arrange, therefore, for special initial configurations that
are invariant under some combination of the above symmetries and be
sure that these conditions will be preserved in time.

We shall choose the configuration to be a replication of $N$ real
coordinates over an infinite complex lattice with periods $\bc_1$ and
$\bc_2$. The $N$ particles effectively live on the real coordinate of a
complex torus. The total number of particles on the covering space
is, thus, infinite and we can parametrize them with the triplet
of indices $(i;m,n)$, where $i=1, \dots N$ labels the particles
in each cell and $m,n \in {\bf Z}$ parametrize the cell.
Shifts in $m$ and $n$ are elements of the (infinite) permutation
symmetry of the system. The kinematical variables are chosen to obey
\begin{eqnarray}
&x_{i;m+1,n} = x_{i;m,n} +\bc_1 ~,~~~ p_{i;m+1,n} = p_{i;m,n} \\
&x_{i;m,n+1} = x_{i;m,n} +\bc_2 ~,~~~ p_{i;m,n+1} = p_{i;m,n} 
\end{eqnarray}
ensuring that we are dealing with lattice copies. This means
\be
x_{i;m,n} = x_i + m\bc_1 + n\bc_2 ~,~~~ p_{i;m,n} = p_i
\label{xpsol}
\ee
The above conditions,
being an invariance under combined permutations and translations,
are dynamically preserved. To also preserve the condition that 
$x_i$ are real, we should impose invariance under
the imaginary parity transformation $x \to x^*$. Choosing $\bc_1$ real,
the only possibilities for $\bc_2$ are
\be
\bc_2 +\bc_2^* =0 ~~~{\rm or}~~~ \bc_2 +\bc_2^* = \bc_1
\ee
The first choice ($\bc_2$ imaginary) leads to an orthogonal lattice,
while the second choice leads to a rhombic lattice.

To ensure full preservation of the lattice structure, we should also
impose appropriate periodic conditions for the spins. In this, we
can take advantage of the global spin symmetry of the system and impose
\begin{eqnarray}
S_{i;m+1,n} &=& U S_{i;m,n} U^{-1} \nonumber \\
S_{i;m,n+1} &=& V S_{i;m,n} V^{-1}
\end{eqnarray}
with $U$ and $V$ two constant matrices.
That is, spins can pick up $U(\n)$ transformations as they
move around the cycles $\bc_1$ or $\bc_2$. Consistency requires that 
$S_{i;m+1,n+1}$ be uniquely determined irrespective of the order 
of increase of the indices $m$ and $n$. That is,
\be
UV S_{i;m,n} V^{-1} U^{-1} =
VU S_{i;m,n} U^{-1} V^{-1} 
\ee
which implies
\be
[ U^{-1} V^{-1} UV, S_{i;m,n} ] =0
\ee
For this to hold for all generic $S_i$ we must require 
$U^{-1} V^{-1} UV \equiv \omega$ to be proportional to the identity matrix.
Clearly $\omega$ satisfies $\det(\omega) = \omega^\n =1$, so we obtain
\be
UV = \omega VU ~,~~~ \omega = e^{i2\pi\frac{\nu}{\n}}
\label{UVcom}
\ee
with $\nu$ an integer $0\leq \nu <\n$. $U$ and $V$ then satisfy 
Weyl's braiding condition which characterizes a noncommutative 
(`quantum') torus \cite{Weyl}.

The spin matrices $S_{i;m,n}$ are now expressed as
\be
S_{i;m,n} = U^m V^n S_i V^{-n} U^{-m} = 
V^n U^m S_i U^{-m} V^{-n}
\label{Ssol}
\ee
Inserting the expressions (\ref{xpsol},\ref{Ssol}) in the hamiltonian
(\ref{Hcal}) we obtain the hamiltonian of the reduced system.
As usual, the resulting hamiltonian has an infinite factor,
due to the summation of the hamiltonians of the infinitely many
identical cells over the complex plane. Dropping this trivial infinity, 
the reduced hamiltonian includes the kinetic terms of the fundamental
cell and the interaction potential of particles in this cell with
all other particles in all cells:
\be
H = \sum_{i=1}^{N} \half p_i^2 + \half \sum_{i,j} 
\sum_{m,n=-\infty}^\infty {\tr( U^m V^n S_i V^{-n} U^{-m} S_j ) 
\over (x_{ij} +m\bc_1+n\bc_2)^2}
\label{Hmn}
\ee
where we adopted the notation $x_{ij} = x_i - x_j$.

\section{Noncommutative spin interaction potentials}

To proceed, we must identify the possible forms of $U,V$. We need the
irreducible representations of the relation (\ref{UVcom}). 
Call $k$ the greatest common divisor of $\nu$
and $\n$. Then $\n = k\m$ and $\nu = k\mu$, for relatively prime $\m,\mu$. 
The irreducible representations for $U,V$ are $\m$-dimensional
`clock' and `shift' matrices. By a global $U(\n)$ spin transformation
we can diagonalize either of $U,V$. Choosing $U$ diagonal, the general
form of $U$ and $V$ will be the direct sum of $k$ of the above irreducible
representations:
\be
U = diag\{e^{i\phi_0} , \dots e^{i\phi_{k-1}} \} \otimes u ~,~~~
V = diag\{e^{i\theta_0} , \dots e^{i\theta_{k-1}} \} \otimes v
\label{UVsol}
\ee
where $\phi_q, \theta_q$ are arbitrary phases, determining the
Casimirs $U^\m$ and $V^\m$, and $u,v$ are the $\m$-dimensional clock
and shift matrices
\be
u_{\alpha \beta} = \omega^{\alpha} \, \delta_{\alpha \beta} ~,~~~ 
v_{\alpha \beta} = \delta_{\alpha+1,\beta}~(mod~\m)
~,~~~ \alpha,\beta=0,\dots \m-1
\ee
So the acceptable $U$ and $V$ depend on $2k$ arbitrary parameters.

To take advantage of the form (\ref{UVsol}) for $U,V$ we partition
$S_i$ into $k^2$ blocks of dimension $\m \times \m$ each by using the
double index notation
\be
(S_i )_{ab} = (S_i )_{\alpha \beta}^{pq}~,~~~ 
a = p\m+\alpha~,~~b=q\m+\beta 
\ee
The $U(\n )$ Poisson brackets in this notation are
\be 
\{ (S_i )_{\alpha \beta}^{pq} , (S_j )_{\gamma \delta}^{rs} \} 
= -i \delta_{ij} \left[ 
(S_i)_{\alpha \delta}^{ps} \, \delta_{\gamma \beta} \, \delta_{rq} 
- \delta_{\alpha \delta} \, \delta_{rq} \, (S_i)_{\gamma \beta}^{rq} \right]
\label{SPB}
\ee
The $m,n$-sums that appear in (\ref{Hmn}) then 
become
\be
\sum_{m,n;\alpha,\beta;p,q} (S_i)_{\alpha+n,\beta+n}^{pq} \,
(S_j)_{\beta,\alpha}^{qp} \,
{e^{-im\phi_{pq} -in\theta_{pq}} \, \omega^{ m (\alpha-\beta)}
\over (x_{ij} +m\bc_1+n\bc_2)^2}
\label{pot}
\ee
where the term $m=n=0$ is omitted if $i=j$. 

The above gives a potential interaction between particles $i$ and $j$
in the form of a modular function in $x_{ij}$ which depends on the
spin components of particles $i$ and $j$. To make the noncommutative
character of the spin interaction manifest, we perform a change
of basis in the spin states and define
\be
({\tilde S}_i )_{\alpha \beta}^{pq} = \sum_\sigma \omega^{\left( \sigma 
+ \frac{\alpha}{2}\right)\beta} \, (S_i )_{\alpha+\sigma,\sigma}^{pq}
\label{Stil}
\ee
This is essentially a discrete Fourier transform in the sum
of the $\alpha,\beta$ indices of $S_{\alpha \beta}^{pq}$.
(Note that, for $\m$ odd, ${\tilde S}_{\alpha \beta}^{pq}$ is
actually antiperiodic in the index $\alpha$ if $\beta$ is odd,
and vice versa. Although we could have defined a properly periodic
matrix, we prefer this slight inconvenience in order to make the
ensuing formulae more symmetric.)
In fact, it will be convenient to assemble the double indices
$(\alpha,\beta)$ and $(m,n)$ into vectors $\vec \alpha$ and
$\vec m$. Similarly, we define $\vec \bc = (\bc_1 , \bc_2 )$
and ${\vec \phi}_p = (\phi_p , \theta_p )$.

The Poisson brackets of the ${\tilde S}_i$ are found from (\ref{SPB})
\be
\{ ({\tilde S}_i )_{\vec \alpha}^{pq} , 
({\tilde S}_j )_{\vec \beta}^{rs} \} = 
i \delta_{ij} \left[ \omega^{\frac{{\vec \alpha} \times {\vec \beta}}{2}}
\, \delta_{ps} \, ({\tilde S}_i )_{{\vec \alpha}+{\vec \beta}}^{rq}
- \omega^{-\frac{{\vec \alpha} \times {\vec \beta}}{2}} \,
\delta_{rq} \, ({\tilde S}_i )_{{\vec \alpha}+{\vec \beta}}^{ps}
\right]
\ee
This is a structure extending the Moyal (star-commutator) algebra, 
the exponent 
of $\omega$ being the cross product of the discrete `momenta' 
$\vec \alpha$ and $\vec \beta$. For $(rs)=(pq)$, in particular,
it becomes the torus Fourier transform of the Moyal bracket
\be
\{ ({\tilde S}_i )_{\vec \alpha}^{pq} , 
({\tilde S}_j )_{\vec \beta}^{pq} \}
= i \delta_{ij} (\omega^\half - \omega^{-\half} ) 
\left[ {\vec \alpha} \times {\vec \beta} \right]_\omega
({\tilde S}_i )_{{\vec \alpha}+{\vec \beta}}^{pq}
\ee
where 
\be
[x]_\omega = \frac{\omega^\frac{x}{2} - \omega^{-\frac{x}{2}}}
{\omega^\half - \omega^{-\half}}
\ee
is the $\omega$-deformation of $x$. This is the so-called
trigonometric algebra with periodic discrete indices \cite{DZ}.

Finally, by inverting (\ref{Stil}) and substituting in (\ref{pot}),
the potential energy $W$ in terms of the ${\tilde S}_i$ acquires the form
\be
W = \sum_{i,j} \sum_{{\vec \alpha};p,q} 
({\tilde S}_i )_{\vec \alpha}^{pq} ~ ({\tilde S}_j )_{-\vec \alpha}^{qp}
~ W_{\vec \alpha}^{pq} ( x_{ij} )
\label{Vfin}
\ee
The above includes two-body interactions, for $i\neq j$, as well as 
spin self-couplings, for $i=j$, arising from the interaction of each 
particle with its own images in different cells. The two-body potential 
$W_{\vec \alpha}^{pq} (x)$ is
\be
W_{\vec \alpha}^{pq} (x) = \frac{1}{\m} \sum_{\vec m}
\frac{ \omega^{{\vec \alpha} \times {\vec m}} \, e^{i {\vec \phi}_{pq} 
\cdot {\vec m}}}{(x + {\vec \bc} \cdot {\vec m} )^2}
\label{Vtwo}
\ee
while the spin self-coupling ${\tilde W}_{\vec \alpha}^{pq}$ is
\be
{\tilde W}_{\vec \alpha}^{pq} = \frac{1}{\m} \sum_{{\vec m} \neq {\vec 0}}
\frac{ \omega^{{\vec \alpha} \times {\vec m}} \, e^{i {\vec \phi}_{pq}
\cdot {\vec m}}}{({\vec \bc} \cdot {\vec m} )^2}
\label{self}
\ee 

If the above potentials were independent of the $U(\n)$ indices
$\vec \alpha$ and $p,q$, the sum over $U(\n)$ indices in the potential
energy expression (\ref{Vfin}) would simply be a $U(\n)$ trace 
and would give the $U(\n)$-invariant coupling between
the spins of particles $i$ and $j$ multiplying the standard Weierstrass
potential of the elliptic Calogero model. In the present case, however,
the above potential is spin-dependent and breaks $U(\n)$ invariance,
introducing a star-product twist in the indices $\vec \alpha$ 
and phase shifts ${\vec \phi}_p$ in the indices $p,q$.
Generically, the $U(\n)$ invariance of the original model is broken
down to an abelian $U(1)^k$, amounting to the transformation
\be
(S_i )_{\alpha \beta}^{pq} \to e^{i\varphi_p} \,
(S_i )_{\alpha \beta}^{pq} ~ e^{-i\varphi_q}
\ee
If ${\vec \phi}_p$ are equal for $k'$ values of $p$, the remaining 
symmetry $U(1)^{k'}$ is enhanced to $U( k' )$, corresponding 
to mixing the corresponding $p$-components.

The case $\omega=1$, ${\vec \phi}_p = 0$
reduces to the standard spin-elliptic Calogero-Moser model. The
case $\m = 1$ (and thus $\omega = 1$) reproduces the 
$U(\n)$-noninvariant model introduced in \cite{APred}. The general
case with $\omega \neq 1$ is a new classical integrable model
of the spin-Calogero type with a spin-dependent potential which
is a modular function of the two-body distance $x_{ij}$. 

The sums appearing in (\ref{Vtwo}) and (\ref{self})
could in principle have ambiguities due to the logarithmic divergence
of the summation over the radial coordinate on the complex plane.
This is, indeed, the case for the standard Weierstrass function and
a specific prescription is needed to regularize it. Different prescriptions
lead to different additive constants in the result. In our case,
however, the presence of the extra phases renders the sums convergent
and there is no regularization ambiguity.

The potentials can be expressed in terms of theta-functions.
$W_{\vec \alpha}^{pq} (x)$ is a modular function 
on the complex torus $(\bc_1 ,\bc_2 )$ with quasiperiodicity
\begin{eqnarray}
W_{\vec \alpha}^{pq} (x+\bc_1 ) &=& e^{-{i\phi_{pq} + \frac{2\pi \mu}
{m} \alpha_2}} ~ W_{\vec \alpha}^{pq} (x) \nonumber \\
W_{\vec \alpha}^{pq} (x+\bc_2 ) &=& e^{-{i\theta_{pq} - \frac{2\pi \mu}
{m} \alpha_1}} ~ W_{\vec \alpha}^{pq} (x) 
\end{eqnarray}
It has a double pole at $x=0$, with principal part
\be
W_{\vec \alpha}^{pq} (x) = \frac{1}{\m x^2} + O( x^0 )
\ee
and no other poles in each cell. These properties uniquely define
$W_{\vec \alpha}^{pq} (x)$ and allow for an expression in terms of 
theta-functions. We put
\be
W_{\vec \alpha}^{pq} (x) = A ~ \omega^{-i\frac{x}{\bc_1}} ~
e^{-{i\frac{x}{\bc_1}\phi_{pq}}} ~ \frac{
\tha \left( \frac{\pi}{\bc_1} (x-Q_1) \right)
\tha \left( \frac{\pi}{\bc_1} (x-Q_2)\right)}
{\tha \left( \frac{\pi}{\bc_1} x \right)^2}
\label{VAQQ}
\ee
where $Q_{1,2}$ are the as yet unknown zeros of $W_{\vec \alpha}^{pq} (x)$
and the theta-functions appearing above have complex period
$T = \bc_2 /\bc_1$. This has the right quasiperiodicity under
$x \to x+\bc_1$. In order to also have the right quasiperiodicity under
$x \to x+\bc_2$, $Q_{1,2}$ must satisfy
\be
Q_1 + Q_2 = \frac{{\vec \phi_{ab}} \times {\vec \bc}}{2\pi}
+ \frac{\mu}{\m} {\vec \alpha} \cdot {\vec \bc}
\label{QQ}
\ee
and to have the right behavior around $x=0$ we must further have
\be
\frac{ \tha' \left( \frac{\pi}{\bc_1} Q_1 \right) }
{ \tha \left( \frac{\pi}{\bc_1} Q_1 \right) } +
\frac{ \tha' \left( \frac{\pi}{\bc_1} Q_2 \right) }
{ \tha \left( \frac{\pi}{\bc_1} Q_2 \right) }
= -i\frac{\phi_{ab}}{\pi} -2i \frac{\mu}{\m} \alpha_1
\label{QQp}
\ee
\be
A = \frac{\pi^2 \, \tha' (0)^2}{\m \, \bc_1^2 \,
\tha \left( \frac{\pi Q_1}{\bc_1} \right)
\tha \left( \frac{\pi Q_2}{\bc_1} \right)} 
\label{AQQ} 
\ee
The equations (\ref{QQ}) and (\ref{QQp}) above determine
$Q_1$ and $Q_2$, while (\ref{AQQ}) in turn determines $A$. 
The self-coupling ${\tilde W}_{\vec \alpha}^{pq}$ can then
be extracted from $W_{\vec \alpha}^{pq} (x)$ as
\be
{\tilde W}_{\vec \alpha}^{pq} = \lim_{x \to 0} 
\left( W_{\vec \alpha}^{pq} (x) - \frac{1}{\m\, x^2} \right)
\label{Vtil}
\ee

The sums appearing in (\ref{Vtwo}) and (\ref{self}) are in general
convergent, due to the presence of the phases.
For $\omega =1$, however, the phases are absent and terms
with $p=q$ have an additive ambiguity due to the need for 
regularization for the expression (\ref{Vtwo}).
In the theta-function expression this manifests in the fact that
the equations for $Q_{1,2}$ (\ref{QQ},\ref{QQp})
are satisfied for {\it any} $Q_1 = -Q_2$.
By applying the addition formula
\be
\tha (x+Q) \tha (x-Q) \thh (0)^2 =  \tha (x)^2 \thh (Q)^2
- \thh (x)^2 \tha (Q)^2
\ee
this is seen indeed to amount to an arbitrary additive constant 
to the expression for $W^{pp} (x)$. The same holds for terms
$p,q$ for which ${\vec \phi}_{pq} = 0$. Such arbitrariness,
however, corresponds to trivial redefinitions of the model
by addition of constants of motion. This is explained in
\cite{APred} and will not be elaborated here.

\section{Conclusions and open questions}

In conclusion, we identified an integrable generalization of the 
elliptic spin model which breaks the spin $U(n)$ invariance
and promotes the potential to a modular function introducing
noncommutative spin twists. 
(Quantum generalizations of elliptic spin models have
appeared recently, but with $U(\n)$ invariant interactions \cite{FGGRZ}.)

There are clearly many issues that deserve further study.
The conserved quantities and Lax matrix of this model can, in 
principle, be obtained from the corresponding quantities of the 
unreduced model; this was done, e.g., in \cite{APred} for
a specific case. A derivation in the present case would be useful.
Further, the modular potential involves implicitly defined $Q_{1,2}$; 
a more explicit and symmetric expression would be desirable. 

The properties of the spin interaction should also be clarified.
In particular, it would be interesting to see if some deformation 
of $U(\m)$ can be identified as a dynamical symmetry.

Finally, the quantum mechanical extension of the model
is, perhaps, the most interesting and pressing question. 
This is not trivially accessible by the method used here since,
in general, the constraints implied by the phase space restrictions 
are second class and we cannot carry over the solution of the 
unrestricted quantum model and apply the constraints as operator
relations on the Hilbert space. The above issues remain interesting
topics for future investigation.

{\it Acknowledgement:} I would like to thank C. Zachos for useful comments
on the manuscript.

\end{document}